\documentclass[sigconf, language=french,language=german, language=spanish, language=english]{acmart}
\AtBeginDocument{%
  \providecommand\BibTeX{{%
    \normalfont B\kern-0.5em{\scshape i\kern-0.25em b}\kern-0.8em\TeX}}}
\setcopyright{none}
\usepackage{tikz}
\usepackage{multirow}

\usetikzlibrary{shapes,arrows,positioning}
\usepackage{graphicx}
\usepackage{caption}
\usepackage{subcaption}

\newcommand{\name}{SpacePulse}

\usepackage{amsmath}
\usepackage{qcircuit}

\renewcommand\footnotetextcopyrightpermission[1]{} 
\begin{document}

\title{SpacePulse: Combining Parameterized Pulses and Contextual Subspace for More Practical VQE}
\author{Zhiding Liang*}
\affiliation{%
  \institution{University of Notre Dame}
  \city{Notre Dame}
  \state{IN}
  \country{USA}
}

\author{Zhixin Song*}
\affiliation{%
  \institution{Georgia Institute of Technology}
  \city{Atlanta}
  \state{GA}
  \country{USA}}
  
\author{Jinglei Cheng*}
\affiliation{%
  \institution{Purdue University}
  \city{West Lafayette}
  \state{IN}
  \country{USA}
}

\author{Hang Ren*}
\affiliation{%
  \institution{University of California, Berkeley}
  \city{Berkeley}
  \state{CA}
  \country{USA}}
  
\author{Tianyi Hao*}
\affiliation{%
  \institution{University of Wisconsin-Madison
}
  \city{Madison}
  \state{WI}
  \country{USA}}
  
\author{Rui Yang*}
\affiliation{%
 \institution{Peking University}
 \city{Beijing}
 \country{China}}

\author{Yiyu Shi}
\affiliation{%
  \institution{University of Notre Dame}
  \city{Notre Dame}
  \state{IN}
  \country{USA}
}

\author{Tongyang Li}
\affiliation{%
 \institution{Peking University}
 \city{Beijing}
 \country{China}}

\thanks{*These authors contributed to the work equally.\\
Corresponding authors: zliang5@nd.edu}

\begin{abstract}

In this paper, we explore the integration of parameterized quantum pulses with the contextual subspace method. 
The advent of parameterized quantum pulses marks a transition from traditional quantum gates to a more flexible and efficient approach to quantum computing. 
Working with pulses allows us to potentially access areas of the Hilbert space that are inaccessible with a CNOT-based circuit decomposition.  
Compared to solving the complete Hamiltonian via the traditional Variational Quantum Eigensolver (VQE), the computation of the contextual correction generally requires fewer qubits and measurements, thus improving computational efficiency. 
Plus a Pauli grouping strategy, our framework, \name, can minimize the quantum resource cost for the VQE and enhance the potential for processing larger molecular structures.

\end{abstract}

\maketitle
\pagestyle{plain}
\section{Introduction}

Quantum computing holds the promise of tackling intricate problems beyond classical computers' capabilities~\cite{Preskill2018NISQ, wang2022quest,li2021co}. 
Quantum computers are built on qubits, which can exist in multiple states simultaneously through quantum superposition. 
This unique property enables parallel processing, delivering significant speed advantages for specific tasks~\cite{daley2022practical,huang2022quantum,riste2012initialization,bravyi2018quantum}. 
However, realizing this potential presents challenges, particularly during the Noisy Intermediate-Scale Quantum (NISQ) era~\cite{wang2022quantumnas, wang2022qoc,zhan2022transmitter}. 
This era is characterized by short coherence times, limited qubit connectivity, and high noise levels~\cite{liang2022pan, wang2022quest, qi2023theoretical,mckinney2023parallel,li2022paulihedral,ding2020systematic}. 
Short coherence times on the order of microseconds, restrict the size of quantum circuits, while connectivity limitations prevent efficient operations among qubits. 
Additionally, interactions with the environment and qubit imperfections contribute to qubits' high noise levels~\cite{cai2022quantum, cheng2020accqoc,chu2023qtrojan,su2017fast,smith2023clifford,li2022design,xu2023systems,das2023imitation,das2022afs}.

To address these challenges, variational quantum algorithms~\cite{cerezo2021variational} have gained prominence, with the extensively studied Variational Quantum Eigensolver (VQE)~\cite{kandala2017hardware,wu2021towards} being a notable example. 
VQE leverages shallow, parameterized circuits to estimate energy or expectation values. 
A classical optimizer updates the circuit parameters before the quantum circuit is instantiated. 
Despite its potential, VQE encounters its own set of challenges~\cite{kandala2017hardware}. 
In the context of solving electronic structure problems, a crucial task in quantum chemistry, the molecular Hamiltonian initially exists in fermionic format and is subsequently mapped to the qubit format. 
This mapping involves expressing the molecular Hamiltonian as a weighted sum of Pauli operators, and the number of Pauli operators determines the quantum circuits that need to be constructed and executed. 
However, it is possible to reduce the number of Pauli operators through techniques such as tapering~\cite{fedorov2022vqe}, contextual subspace~\cite{kirby2021contextual}, and grouping~\cite{tilly2022variational}, which, in turn, decreases the overall number of circuits and their sizes. 

To enhance the practicality of VQE, another promising approach is the utilization of variational pulse circuits~\cite{liang2022variational,meitei2021gate,liang2023hybrid}. 
Within variational pulse circuits, the parameters include pulse amplitudes, angles, and durations~\cite{egger2023study, meitei2021gate,liang2023towards, kottmann2023evaluating,puzzuoli2023qiskit}. 
This shift from gate circuits to pulse circuits can lead to reduced circuit latency.
In this paper, we focus on the integration of contextual subspace~\cite{kirby2019contextuality} with pulse circuits. 
The primary objective of this integration is to minimize the overhead in terms of quantum resources. Specifically, our work has the following contributions:

\begin{itemize}
    \item The introduction of a framework designed to minimize quantum resource consumption, \name, which through several steps effectively reduces the number of qubits, the quantity of measurements, and the duration of the quantum circuit. This approach is shown to significantly streamline the quantum computation process.
    \item The novel conceptualization and successful implementation of a contextual subspace model that utilizes a pulse ansatz. This model has been experimentally validated on actual quantum hardware, demonstrating its advantages over the gate ansatz version of the contextual subspace model. 
\end{itemize}
\begin{figure*}[t]
\centering
\includegraphics[width=\linewidth]{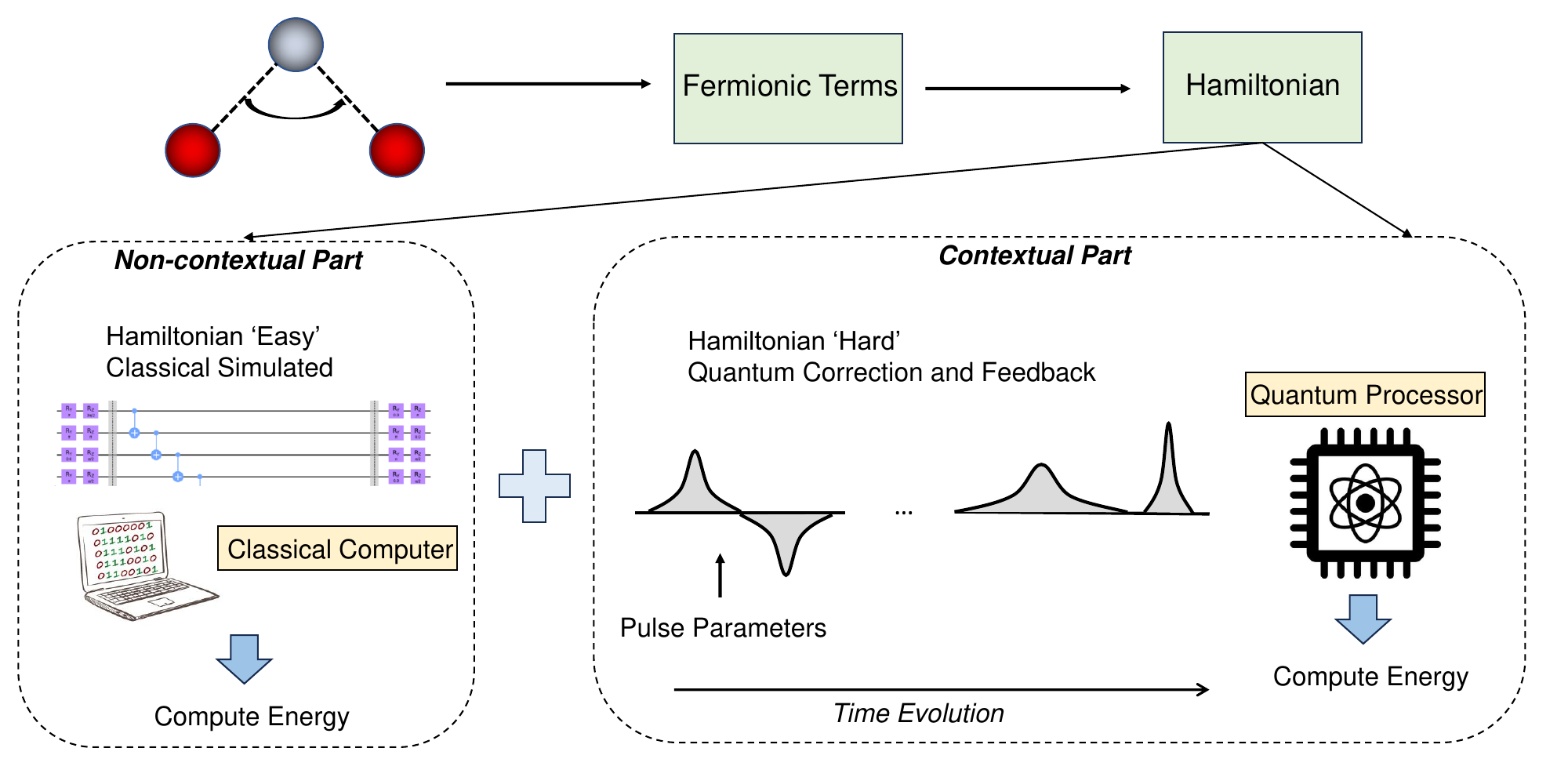}
\caption{
Overview of the \name\ , a combining framework of the contextual subspace method and the parameterized pulse-based ansatz. 
}

\vspace{-4mm}
\label{teaser}
\end{figure*}
\section{Background and Motivation}
\subsection{Variational Quantum Eigensolver}

The Variational Quantum Eigensolver (VQE) is designed to work with the capabilities of today's noisy quantum devices~\cite{fedorov2022vqe}. 
Unlike classical-quantum algorithms that need fault-tolerant machines, VQE is made to handle the less-than-perfect conditions of current NISQ technology. 
The core of VQE involves representing a system's Hamiltonian using weighted Pauli operators, which are then transformed into quantum circuits for computation. 
A crucial part of this process is the ansatz, a quantum circuit with adjustable parameters to prepare the ground state of the Hamiltonian~\cite{weaving2023stabilizer}. 
By measuring this prepared state, we can determine an upper bound for the system's ground state energy, a crucial value in quantum physics and chemistry. 
The strength of VQE lies in its iterative nature, where a combination of quantum state preparation and classical optimization continually updates the ansatz parameters. 
This iterative optimization minimizes the expectation value of the Hamiltonian, getting closer to the system’s lowest energy configuration with each iteration.

\subsection{Contextual Subspace Method}

The Hamiltonian of interest, denoted as $H$, is divided into two components: the noncontextual part $H_{nc}$, which contributes classically solvable energy to the system's Hamiltonian, and the contextual part $H_{c}$, whose energy contribution is computed using a VQE on quantum processors~\cite{ralli2023unitary, kirby2023contextuality, ralli2023practical}. 
This serves as a quantum correction to the energy optimized classically. The VQE is constrained such that the search space aligns with the noncontextual ground state. This method is termed the contextual subspace variational quantum eigensolver (CSVQE)~\cite{kirby2021contextual}. 
This approach is advantageous in terms of experimental resource utilization, requiring fewer qubits and measurements. 
Following this approach, the Hamiltonian can be represented as:
\begin{equation}
H = H_{nc} + H_c.
\end{equation}

Specifically, let the Pauli terms of $ H_{nc}, H_c, H$ be $S_{nc}, S_c, S$, respectively, where $S = S_{nc} \cup S_c$. Noncontextuality implies the possibility of assigning definite values to all Pauli terms $S_{nc}$ simultaneously without contradiction \cite{kirby2019contextuality}. 
For instance, the Pauli operator set $\{IZ, XI\}$ is noncontextual as its elements commute, allowing for concurrent measurements and joint value assignments. 
In contrast, contextuality, a unique quantum characteristic, refers to the feature that such joint assignments are not feasible, leading to inconsistencies in value inference. 
For example, the set $\{IZ, XI, IX, ZI\}$ is contextual; assigning values $\{v_{IZ},v_{XI}, v_{IX}, v_{ZI}\} $ and inferring the value of $v_{YY}$ can lead to contradictory results: We can either calculate $ YY = (ZZ = ZI \times IZ) \times (XX = XI \times IX)$ and derive that $v_{YY} = v_{IZ}v_{XI}v_{IX}v_{ZI}$ or through $ YY = -(ZX = ZI \times IX) \times (XZ = XI \times IZ)$ to get $v_{YY} = -v_{IZ}v_{XI}v_{IX}v_{ZI}$. 
Obviously, these two inferences contradict, which indicates the existence of contextuality. 
The method for partitioning the Hamiltonian into contextual and noncontextual parts is discussed in \cite{kirby2020classical}.

For the noncontextual portion, it can be proved that any noncontextual set $S_{nc}$ takes the form:
\begin{equation}
S_{nc} = Z \cup C_1 \cup C_2 \cup \dots \cup C_N.
\end{equation}

Here, any element $A \in Z$ commutes with any $B \in S_{nc}$, and the $C_i$ pairwise anticommute. 
Given that $S_{nc}$ can have simultaneous value assignments, they share the same eigenspace. 
The eigenstates of $H_{nc}$ are characterized by the probability distributions over this space. 
The expectations of Hamiltonian terms can be recovered by inference on:
\begin{equation}
    G  \cup \{A_1\} \cup \{A_2\} \cup \dots \cup \{A_N\},
\end{equation}
where $G$ is the generator set for $Z$ and $A_i \in C_i$. 
For these eigenstates to be physical, i.e., for the noncontextual states to be valid quantum states with correct probability distributions, the value assignment must satisfy:
\begin{equation}
    \langle G_j\rangle=q_j= \pm 1, \quad\left\langle A_i\right\rangle=r_i, \quad|\vec{r}|=1
\end{equation}

The problem of finding the ground energy of $H_{nc}$ thus translates into minimizing the expectation value by variational search within the allowed parameter space $(\vec{q}, \vec{r})$. 
It can be proved that this space suffices to generate all potential expectation values of the Hamiltonian. 
It now becomes a constrained optimization problem solvable using classical computation. 
With this method, the noncontextual Hamiltonian can be solved efficiently using classical computers.

Denoting $\psi_{(\vec{q}, \vec{r})}$ as the noncontextual ground state with energy $E_{nc}$, we thus have
\begin{equation}\label{E_c}
    E_c =\langle \psi_{(\vec{q}, \vec{r})} | H_c |\psi_{(\vec{q}, \vec{r})}\rangle.
\end{equation}

The subsequent step involves addressing the contextual component. 
The noncontextual states are in the subspaces of quantum states stabilized by the operators  $G\vec{q}$ and
$\vec{A}$. 
This subspace is referred to as the contextual subspace, within which the search for the contextual ground state should be confined.

A set of stabilizers $W$ determines the subspace. 
Correspondingly, a projector $Q_W$ can be identified that maps the Hilbert space onto this subspace.

To ensure consistency between the contextual and noncontextual ground states, we constrain the contextual Hamiltonian within the subspace and construct a transformed Hamiltonian:
\begin{equation}
H_{c} \mapsto H_{c }^{W}=Q_{W}^{\dagger} U_{W}^{\dagger} H_{c} U_{W} Q_{W}.
\end{equation}
$H_{c }^{W}$ now becomes the target Hamiltonian for the VQE. The VQE circuit is structured as:
\begin{equation}
|\psi_{c}(\vec{\theta})\rangle=U_{W}^{\dagger} V(\vec{\theta})|0\rangle^{\otimes n}.
\end{equation}
Here, $V(\vec{\theta})$ represents the parameterized circuit for variational optimization. Using a quantum processor for VQE in the conventional manner allows for determining the energy contribution from the contextual part:
\begin{equation}\label{E_nc}
E_c = \frac{\langle\psi_{c }(\vec{\theta})|Q_{W}^{\dagger} U_{W}^{\dagger} H_{c} U_{W} Q_{\mathcal{W}}| \psi_{c }(\vec{\theta})\rangle}{\langle\psi_{ c}(\vec{\theta})|Q_{W}^{\dagger} Q_{\mathcal{W}}| \psi_{c}(\vec{\theta})\rangle}.
\end{equation}
At the end, the ground energy computed via CSVQE is the sum of contributions from the noncontextual part in Eq.~(\ref{E_c}) and contextual part in Eq.~(\ref{E_nc}):
\begin{equation}
    E_{CSVQE} = E_c + E_{nc}.
\end{equation}

CSVQE has demonstrated potential in quantum chemistry applications~\cite{ravi2022cafqa}. It allows for a significant reduction in the number of qubits required to achieve chemical accuracy, with the number of terms necessary for computing the contextual correction reduced by over an order of magnitude. This indicates that CSVQE is a promising approach for solving practical problems on noisy intermediate-scale quantum devices.

\subsection{Parameterized Pulse-based Ansatz}
The pulse layer is the basic building block for quantum computing in software and systems~\cite{alexander2020qiskit}. 
In this hierarchy, high-level programming designed for gate-level quantum circuits is translated into quantum pulses. 
Quantum pulses are traditionally associated with calibrating basic gates in a quantum computer. 
For example, single-qubit gates fine-tune their angles and amplitudes through Rabi oscillation experiments~\cite{sheldon2016characterizing}, and two-qubit gates are calibrated using Hamiltonian tomography~\cite{sheldon2016procedure}.

Different techniques are proposed to take advantage of quantum pulses. 
For example, quantum optimal control (QOC)~\cite{cheng2020accqoc, werschnik2007quantum} generates pulses based on a prescribed unitary matrix. 
However, QOC possesses considerable computational resource demands.
For quantum pulses, the qubit Hamiltonian is determined by
Eq.~(\ref{pulsesimu}), where $H_d(t)$ is the uncontrollable drift Hamiltonian in real quantum systems, and $H_c(t)$ is the control Hamiltonian that we can modify by adjusting the pulse shapes represented by $c_j(t)$.
\begin{equation}
H(t) = H_d(t) + H_c(t) = H_d(t) + \sum_{j} c_j(t) H_j.
\label{pulsesimu}
\end{equation}

This paper focuses on exploring microwave pulse control in superconducting quantum computers, where key parameters include amplitude, angle, duration, and frequency. 
Adjusting these parameters directly impacts the driving Hamiltonian of the quantum operation, as depicted in Eq.~(\ref{pulseham}), and consequently influences the evolution of the state within the quantum circuit. 
In this equation, \( V_0 \) represents the amplitude, I and Q are ``in phase'' and ``out phase'' controlled by the parameter \( \phi \), which is the angle, \( \omega_q t \) is the qubit frequency, \( \omega_d t \) is the drive frequency, and  \( \Omega \)  is the product of the total capacitance and the charge variable of the circuit.
\begin{equation}
\begin{split}
H_d = & \Omega V_0 s(t) \big( I \sin(\omega_d t) - Q \cos(\omega_d t) \big) \\
      & \times \big( \cos(\omega_q t) \sigma_y - \sin(\omega_q t) \sigma_x \big).
\end{split}
\label{pulseham}
\end{equation}


Working with quantum pulses can potentially explore areas of the Hilbert space that are not easily reachable using traditional CNOT-based circuit decomposition~\cite{egger2023study}. 
This is especially useful for variational algorithms that can take advantage of these otherwise hard-to-reach areas to find more direct paths to solutions, often referred to as ``shortcuts''. 
In the context of NISQ devices operating in noisy environments, controlling at the pulse level is often closer to the Hilbert space than gate-level control for the same search space. 
Directly manipulating quantum pulses also allows us to avoid relying on intermediate gate representations. 

\section{Method}


In our study, we present the \name\  model in the field of VQE. 
This model is specifically designed to tackle the challenges that are currently faced in the early stages of quantum computing. 
These challenges include the issues of noise and limited resources in quantum hardware.
Our model introduces a hybrid computational approach, combining the strengths of classical algorithms with the unique properties of quantum mechanics.
A key feature of the \name\  model is its innovative use of a contextual subspace method coupled with a parameterized pulse ansatz, unlike other contextual subspace models like CSVQE~\cite{kirby2021contextual} and CAFQA~\cite{ravi2022cafqa} focus on gate-level solutions.
This approach significantly reduces the effects of short decoherence times and loosens the requirements for quantum resources.
For a clearer understanding, the \name\  model is depicted in Figure 1. 

\begin{figure*}[t]
\centering
\includegraphics[width=\linewidth]{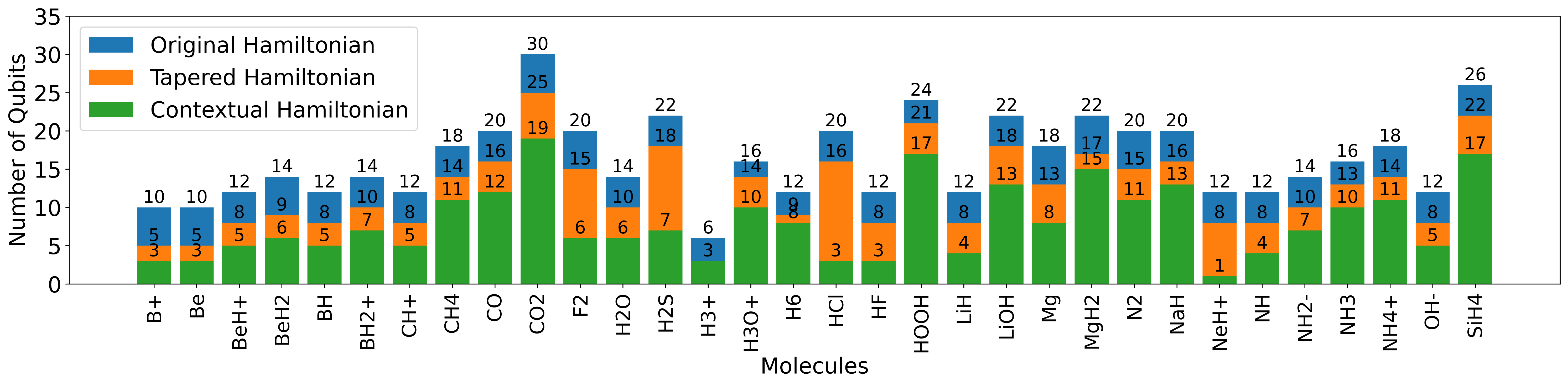}
\caption{
Comparative analysis of qubit requirements for electronic structure calculations across various Hamiltonians and contains the number of qubits for the original, tapered, and contextual Hamiltonians. Each set of bars corresponds to a different molecular system, illustrating the efficacy of tapering and the contextual subspace method in reducing qubit count while aiming for chemical accuracy. The numbers atop the bars indicate the absolute number of qubits for each Hamiltonian form.
}

\vspace{-4mm}
\label{cstapering}
\end{figure*}
\textbf{Tapering Process in the \name\   Model:}
Tapering is the first step in preparing the \name\ model for quantum computing. 
Its main purpose is to reduce the number of qubits. 
The process starts by examining the molecular Hamiltonian to find specific patterns known as Z2 symmetries which are pairs of qubits that are related such that their overall contribution to the state of the system can be predicted from one another.
By identifying these symmetries, tapering can effectively remove qubits that are redundant. 
For every Z2 symmetry found, one qubit can be taken out. 

\textbf{Hybrid Classical-Quantum Framework for Molecular Energy Calculation:} 
The \name\ model integrates quantum pulses with a key innovation: the contextual subspace method. 
This method divides a molecule's Hamiltonian into two segments: ``easy'' and ``hard''. 
The ``easy'' segment can be efficiently processed using classical computers by mapping it onto a stabilizer group. 
This leverages the reliability of classical computing for less complex calculations. 
On the other hand, the ``hard'' segment, which involves more quantum phenomena, is tackled using a hardware-efficient ansatz on quantum hardware. 
Through the hybrid classical-quantum framework, the \name\ model effectively balances classical and quantum computing strengths, improving the efficiency of molecular energy computations.

\textbf{Pauli Grouping Strategy:} 
The \name\ model employs a Pauli Grouping Strategy to further improve the performance. 
This strategy organizes Pauli operators into commuting groups~\cite{gokhale2020optimization,gokhale2019minimizing}. 
Commuting operators can be applied in any sequence without changing the results. 
A straightforward example is that the operators XIXI, IXIX, and IIXX can be simultaneously measured by constructing circuits for XXXX.
By grouping commuting Pauli operators found in the Hamiltonian, the model allows for their simultaneous measurement. 
This approach significantly accelerates the measurement process, reducing the number of measurements needed.

\textbf{Parameterized Pulse Ansatz for Quantum Hardware:} 
The \name\ model uses a parameterized pulse ansatz~\cite{liang2022pan} for the ``hard'' part of the Hamiltonian in quantum hardware. This method is different from traditional gate-level circuits, as it uses pulse-level variational quantum circuits. The main goal is to iteratively update the pulse parameters to efficiently find the correct Hamiltonians that evolves the quantum system from its initial state to the desired ground state. This pulse-level approach shortens the circuit's duration while maintaining accuracy. 

\section{QUALITATIVE ANALYSIS}
\subsection{Less Qubits Usage by Tapering and Contextual Subspace Method}
The Figure. \ref{cstapering} summarizes the results of applying tapering and the contextual subspace method to a set of electronic structure Hamiltonians. We compared the initial Hamiltonian, the Hamiltonian after conical truncation, and the Hamiltonian further processed using the contextual subspace method, reporting the number of quantum bits required to achieve chemical accuracy in each Hamiltonian. We found that conical truncation can reduce the number of quantum bits to varying degrees, and subsequently applying the contextual subspace method can further reduce the number of qubits to our preset threshold, which can be minimal, or even zero. In such cases, only the uncontextual part is computed classically. Our goal is to demonstrate that even though CS-VQE is generally an approximate method, significant qubit reduction can still yield chemical accuracy.

We set this threshold from the number of qubits equal to one, iteratively calculating the molecular energy at the current setting, until the number of qubits after tapering. If chemical accuracy (< FCI energy + 0.0016) is reached during the process, we stop and consider that the current threshold is the minimum number of qubits required for the problem after applying tapering and the contextual subspace method.

\begin{figure}[t]
\centering
\includegraphics[width=\linewidth]{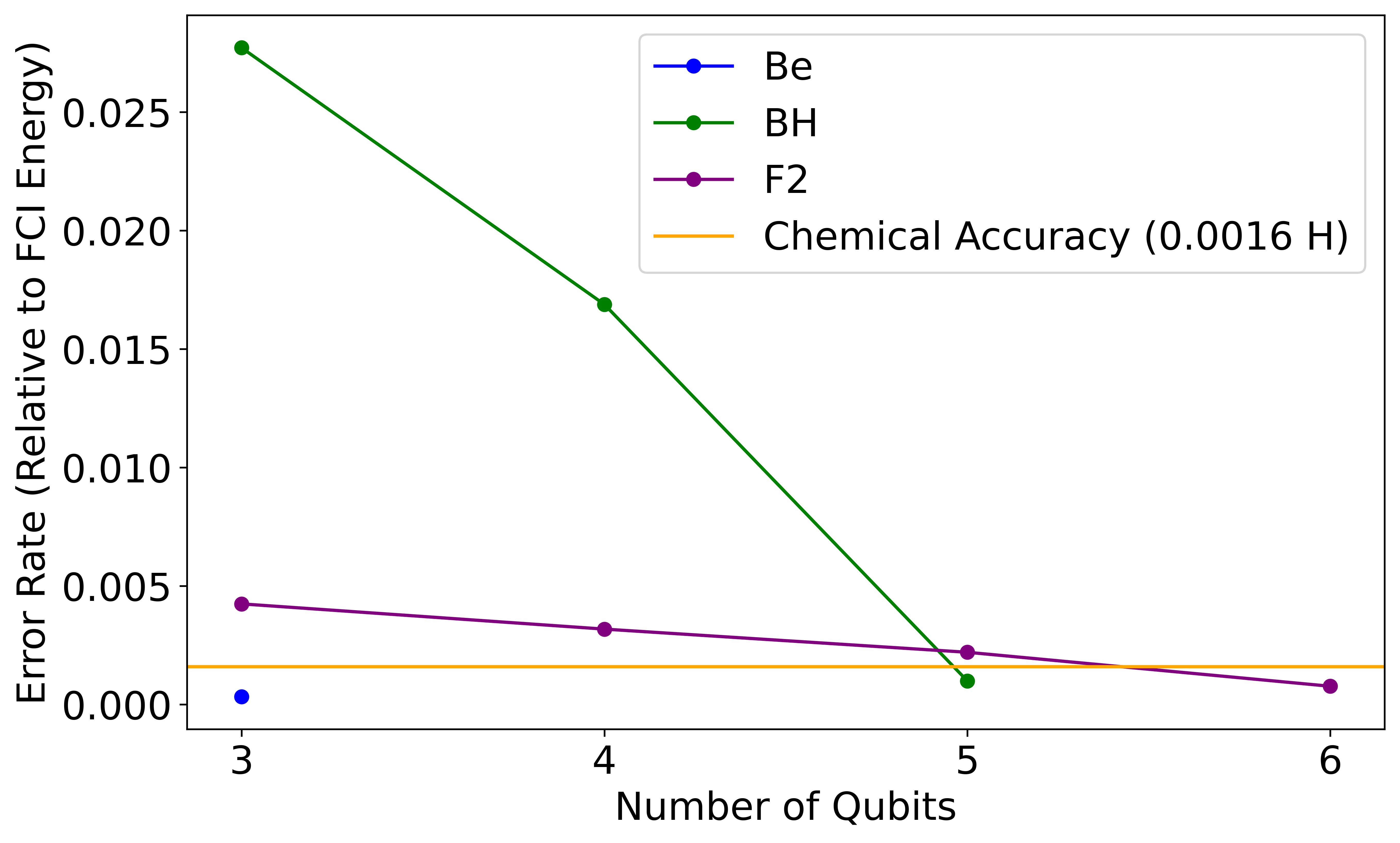}
\caption{
Illustrating the process of approaching chemical accuracy with different numbers of qubits set as thresholds for the contextual subspace method. The graph shows the error rate relative to FCI energy for $Be$, $BH$, and $F_2$ molecules. The red line represents the chemical accuracy threshold of 0.0016 Hartree.
}

\vspace{-4mm}
\label{chemical}
\end{figure}

In the Figure. \ref{chemical}, we show examples of how we selected the minimum number of qubits required by the contextual subspace method for different molecules. We computed the molecular energies after applying both tapering and the contextual subspace method, to ascertain the minimum number of qubits needed to achieve chemical accuracy for different molecules. From the figure, we can see the error rate relative to FCI energy for $Be$, $BH$, and $F_2$ molecules. For $Be$, the chemical accuracy was achieved directly at three qubits, while $BH$ reached it at five qubits and $F2$ at six qubits. Despite reducing the number of qubits by approximately an order of magnitude, the combination of tapering and the contextual subspace method still ensures chemical accuracy in a perfect environment. 

\begin{figure}[t]
\centering
\includegraphics[width=\linewidth]{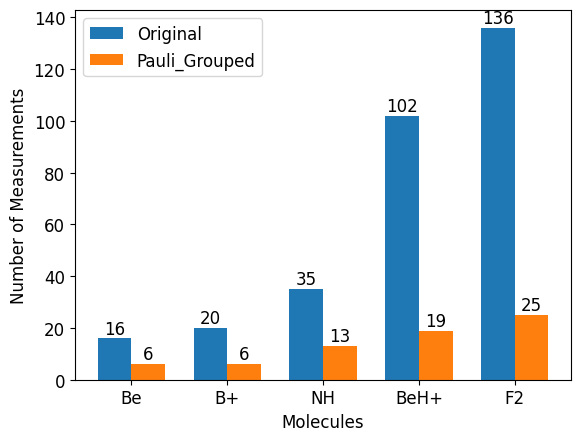}
\caption{
Shows the number of measurements in five different molecular tasks with and without the Pauli grouping strategy. ``Original'' represents the original number of measurements, and ``Pauli\_Grouped'' represents the number of measurements after using the Pauli grouping strategy.
}

\vspace{-4mm}
\label{grouping}
\end{figure}

\subsection{Less Measurement Required by Pauli Grouping Strategy}
Figure \ref{grouping} presents a comprehensive summary of the results obtained from applying the Pauli grouping strategy to a series of electronic structure Hamiltonians. We meticulously report the number of terms required in each Hamiltonian to achieve chemical accuracy in comparison with the original unmodified problem. In our approach, we partitioned the Hamiltonians into sets of commuting groups, a process that assured a reduction in the number of required measurements while preserving the accuracy of the computations. Upon detailed observation, we discovered that for the majority of the chemical molecules discussed, the number of measurements required could be significantly reduced through the Pauli grouping strategy. For example, the number of measurements for small tasks like $Be$ could reduced from 16 to 6, for larger tasks like $F_2$, we observed that the Pauli grouping could reduce over 100 measurements. This reduction was notable even after the qubit count had been substantially decreased through the application of tapering and the contextual subspace method. Specifically, we found that for these molecules, the number of measurement terms necessary could be reduced to approximately one-third of the original count.
\begin{figure}[t]
\centering
\includegraphics[width=\linewidth]{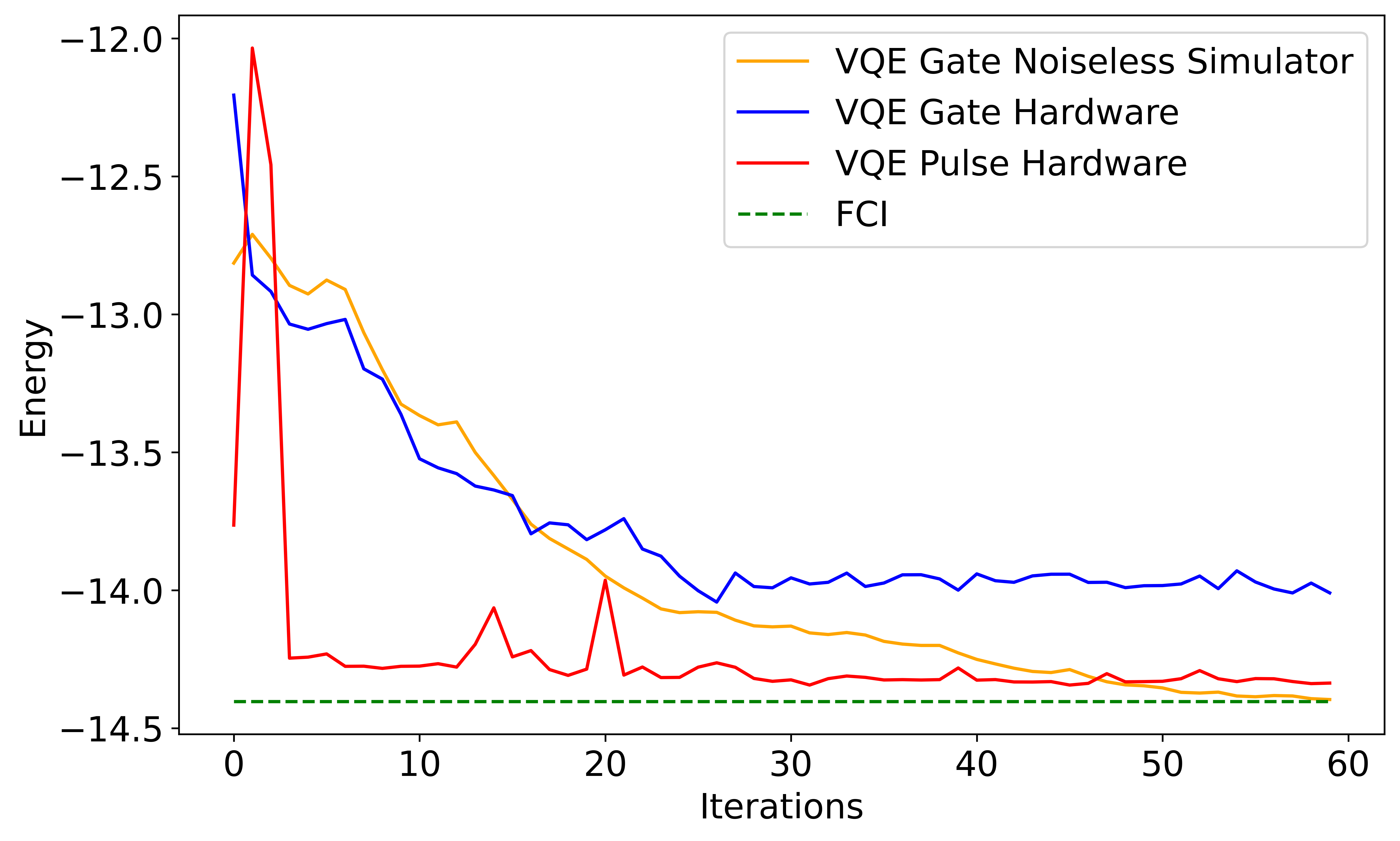}
\caption{
Energy convergence of a $Be$ as a function of VQE iterations. The figure compares the performance of two ansatz methods: a gate-level ansatz and a pulse-level ansatz, both on ibmq\_kolkata (``VQE Gate Hardware'' and ``VQE Pulse Hardware'' in the label of the figure), as well as a gate-level ansatz on a noiseless simulator, the ``VQE Gate Noiseless Simulator'' in the label of the figure. The FCI energy line serves as a benchmark for the ground state energy.
}

\vspace{-4mm}
\label{vqe}
\end{figure}
\begin{table*}[]
\centering
\renewcommand*{\arraystretch}{2}
\setlength{\tabcolsep}{8pt}
\footnotesize
\begin{tabular}{ccccccccc}
\hline
\multirow{2}{*}{\begin{tabular}[c]{@{}c@{}} Molecules \end{tabular}} & \multirow{2}{*}{\begin{tabular}[c]{@{}c@{}}Full \#Qubits\end{tabular}} & \multirow{2}{*}{\begin{tabular}[c]{@{}c@{}}Remaining \#Qubits\end{tabular}} & \multirow{2}{*}{\begin{tabular}[c]{@{}c@{}}Full \#Measurement\end{tabular}} & \multirow{2}{*}{\begin{tabular}[c]{@{}c@{}}Remaining \#Measurement\end{tabular}} & \multicolumn{2}{c}{Gate Ansatz + DD} & \multicolumn{2}{c}{Pulse Ansatz} \\ \cline{6-9} 
                           &                                                                                       &                                                                                      &                                                                                            &                                                                                           & Accuracy    & Duration               & Accuracy        & Duration       \\ \hline
NH                         & 12                                                                                    & 4                                                                                    & 35                                                                                         & 13                                                                                        & 99.59\%     & 6336dt                 & 99.72\%         & 1184dt         \\
BeH+                       & 12                                                                                    & 5                                                                                    & 102                                                                                        & 19                                                                                        & 99.29\%     & 8064dt                 & 99.00\%         & 1328dt         \\
F2                         & 20                                                                                    & 6                                                                                    & 136                                                                                        & 25                                                                                        & 98.61\%     & 11232dt   & 99.38\%         & 1536dt         \\ \hline
\end{tabular}%
\caption{Comparing the results of simulating different molecules on the IBM Quantum computer ``ibmq\_kolkata'' using gate ansatz with dynamic decoupling (DD) and pulse ansatz. The table lists molecules $NH$, $BeH+$, and $F_2$, and provides the accuracy and duration for tasks using gate ansatz with DD and pulse ansatz.}
\label{result}
\vspace{-4mm}
\end{table*}

\subsection{Less Circuit Duration by Construct Pulse Ansatz}
The Figure. \ref{vqe} compares energy estimations for a $Be$ molecule using different ansatz methods on different platforms. It contrasts the use of a gate-level ansatz on a noiseless simulator and real hardware (ibmq\_kolkata), against a pulse-level ansatz on ibmq\_kolkata. The gate ansatz employed is EfficientSU2, a proven hardware-efficient ansatz for VQE tasks with a circuit duration of 4224 dt. The pulse ansatz, however, marks a significant reduction in circuit duration to 768 dt, amounting to an 81.82\% decrease. This reduction is particularly relevant for NISQ machines due to their limited decoherence time, and using a pulse ansatz has the potential to implement more circuit layers within the same decoherence window.

Furthermore, circuit performance is a critical metric of interest, as the aim is not to compromise performance for shorter circuit length. Should such a compromise be necessary, it would prompt a re-evaluation of the tradeoff between saved duration and reduced performance. However, observations indicate that the pulse ansatz achieves higher accuracy for the same task within a much shorter circuit duration. Accuracy was calculated by dividing the estimated energy by the FCI (Full Configuration Interaction) energy, with the pulse ansatz on ibmq\_kolkata reaching an accuracy of 99.7\%. In comparison, the gate ansatz achieves an accuracy of 97.364\% on ibmq\_kolkata, and 99.949\% on a noise-free simulator. These observations suggest that the pulse ansatz we have constructed performs substantially better on molecular problems that have been processed through contextual subspace, tapering, and Pauli grouping, while also offering significant advantages in circuit duration.

\section{Evaluation}

\subsection{Experiment Setup}
Our experimental investigations were carried out on real NISQ devices, specifically utilizing the \texttt{ibmq\_kolkata} from the IBM Quantum systems suite. For our VQE tasks, we selected a diverse set of molecules, including $Be$, $NH$, $BeH+$, and $F_2$, as our test cases. Throughout the optimization process, we employed the \texttt{COBYLA} optimizer, a derivative-free optimizer well-suited for the noisy quantum landscape. We set the cap for iterations at 100 to ensure a thorough exploration of the parameter space while maintaining computational efficiency. In each iteration, we ran the hybrid quantum-classical program 2300 times to accumulate a robust statistical sample for measurement.

Additionally, in light of the observed discrepancy in performance between the gate ansatz and the pulse ansatz in Section 4.3, we integrated a dynamic decoupling strategy for the gate ansatz. This technique was selectively applied to enhance the gate ansatz performance, aiming to counteract the effects of noise and decoherence without interfering with the system's innate dynamics. Through this approach, we sought to isolate and amplify the benefits of the gate ansatz, providing a clear comparative analysis with the pulse ansatz under equivalent conditions.

\subsection{NISQ Device Results}
In this study, we conducted a comprehensive evaluation of three quantum simulation methodologies on the ``ibmq\_kolkata''. As shown in Table \ref{result}, the techniques assessed include the gate ansatz augmented with dynamic decoupling (DD) and the pulse ansatz, applied to the simulation of three molecules: $NH$, $BeH+$, and $F_2$. Our findings indicate that simulation accuracy remains impressively high across all three methods. Specifically, the gate ansatz with DD achieved an accuracy of 99.59\% for NH, 99.29\% for BeH+, and 98.60\% for $F_2$. The pulse ansatz demonstrated a slight increase in accuracy, with 99.72\% for NH, 99.00\% for BeH+, and 99.38\% for $F_2$, suggesting a marginal edge in precision.

In terms of computational duration, measured in discrete time units ``dt'', which equals 0.22ns, we observed notable differences. The gate ansatz with DD recorded durations of 6336dt for $NH$, 8066dt for $BeH+$, and 11232dt for $F_2$. In contrast, the pulse ansatz significantly reduced the circuit duration to 1184dt for $NH$, 1238dt for $BeH+$, and 1536dt for $F_2$. Pulse ansatz shows a huge advantage over the gate ansatz in Sec 4.3, and is still comparable as we implement the DD to the gate ansatz, even though the gate ansatz with DD still shows worse performance than the gate ansatz with even 7.31x circuit duration on $F_2$. We would rather further implement zero noise extrapolation (ZNE) for this task. ZNE is a strong quantum error mitigation (QEM) technique that has been proven to boost the performance of quantum circuits, and we observed the accuracy to be 99.957\%, which is better than the pulse ansatz, but notably, our research is orthogonal to the QEM, ZNE can also implement the pulse ansatz that is proved in~\cite{cheng2023fidelity}. This demonstrates that the pulse ansatz not only maintains high accuracy but also achieves substantial improvements in circuit duration. 

\section{Conclusion}
Our work presents a significant advance in the application of quantum computing to molecular simulations, leveraging the synergy between parameterized quantum pulses and the contextual subspace method. We have demonstrated that through our integrated approach, which we have denominated \name, we can achieve high accuracy in simulating the electronic structure of molecules while utilizing fewer quantum resources. Our experiments on the ``ibmq\_kolkata'' quantum computer have shown that the pulse ansatz not only maintains accuracy but also significantly reduces the computational duration compared to the gate ansatz, even when dynamic decoupling strategies are applied to mitigate noise.

This reduction in resource requirements and increase in computational efficiency are critical for performing quantum simulations on NISQ devices, where resources are scarce and decoherence times are limited. \name's ability to reduce the quantum bit overhead and measurement operations while maintaining simulation fidelity underscores its potential for broader adoption and its suitability for tackling more complex molecular systems in future quantum devices. The study establishes a new benchmark for quantum simulations of molecular systems, pushing the boundaries of what is currently achievable in the NISQ era and opening up pathways for practical quantum computing applications in the field of quantum chemistry.
\section*{Acknowledgment}
We acknowledge the use of IBM Quantum services for this work. And we thank Feng Qian for the valuable discussion about the contextual subspace method.

\bibliographystyle{ACM-Reference-Format}
\bibliography{sample-base}

\appendix









\end{document}